\documentclass[preprint]{elsarticle}
\usepackage{lineno,hyperref}
\usepackage{amsmath,amssymb}
\usepackage{graphicx}
\usepackage[utf8]{inputenc}
\usepackage[english]{babel}
\usepackage[T2A]{fontenc}
\usepackage{comment} 
\modulolinenumbers[10]
\modulolinenumbers[1]
\usepackage{amsthm}
\usepackage{color}
\usepackage{geometry}		
\usepackage{mathtools}
\usepackage{subfig}
\begin{document}
\begin{frontmatter}
\title{Discrete hybrid Izhikevich neuron model: nodal and network behaviours considering electromagnetic flux coupling}

\author{Sishu Shankar Muni}

\ead{s.muni@massey.ac.nz}

\address{School of Fundamental Sciences, Massey University, Palmerston North, New Zealand} 

\author{Karthikeyan Rajagopal}

\ead{karthikeyan.rajagopal@citchennai.net}

\address{Center for Nonlinear Systems, Chennai Institute of Technology,Chennai,Tamil Nadu-600069, India} 

\author{Anitha Karthikeyan}

\ead{mrs.anithakarthikeyan@gmail.com}

\address{{Department of Electronics and Communication Engineering, Prathyusha Engineering College, Thiruvallur, Tamil Nadu-602025, India}} 

\author{Sundaram Arun}

\ead{aruns@prathyusha.edu.in}

\address{Department of Electronics and Communication Engineering, Prathyusha Engineering College, Thiruvallur, Tamil Nadu-602025, India} 

\begin{abstract}
We analyse the dynamics of the improved discretised version of the well known Izhikevich neuron model under the action of external electromagnetic field. It is found that the three-dimensional IZH map shows rich dynamics. With the variation of the electromagnetic field, period-doubling route to chaos in a repeating fashion is observed from the bifurcation diagram. Even the forward and backward continuation bifurcation diagram which do not completely overlap suggests that there is multistability in the system. The phenomenon of bistability (coexistence of periodic and chaotic attractors) is observed. The presence of periodic and chaotic attractor is aided by the maximal Lyapunov exponent diagram. The Lyapunov phase diagram of electromagnetic field and synapses current shows a large parameter region of chaotic and periodic behaviors with the presence of unbounded regions as well. The IZH map shows a plethora of spiking and bursting patterns such as mixed-mode patterns, tonic spiking, phasic spiking, steady spikes, regular spikes, spike bursting, periodic bursting, phasic bursting, chaotic firing etc with the variation of electromagnetic coupling strength and the synapses current. We also investigate the presence of chimera states in a ring-star, ring, star networks of IZH map neurons. Chimera states are found in the case of ring-star and ring network while synchronised clusters were found in the case of star network and are aided by the spatiotemporal plots, space-time plot, recurrence plots. The rich dynamics shown by the discretised IZH map makes it a promising research model to study about neurodynamics. 
\end{abstract}
\end{frontmatter}
\maketitle
\section{Introduction}
It is of utmost importance to study the neuron behavior as it is a major component and the most basic cell of the brain. The study of neuron dynamics is one of the emerging field in nonlinear dynamics. To mimic the neuron behavior and understand its dynamics better, neurons are modelled using systems of ordinary differential equations. To be classified as a neuron model, the model should describe the bursting and spiking features. The model should be computationally easier as well in the sense that it can simulate a large number of neurons within less time. This is essential as in real case, in the brain a large number of neurons are interconnected in a very complicated network. It is with the use of computational techniques, researchers can explore many different and complicated questions of neuron dynamics. The same applies for the study of different neuron models, network of neuron models. Hodgin-Huxley model \cite{HoHu52} was the first neuron model in which neuron exhibits a lot of similar features as that of an actual neuron and was one of the greatest achievements in the field of neuron modelling. It is found to be computationally expensive when implemented on a network of neurons and many extended models were developed from the Hodgkin-Huxley model. Izhikevich also modelled a similar simplified model of a neuron \cite{IZH03}. It is a two dimensional continuous neuron model which is much simpler (in the sense that it has only one nonlinear term in its state equation), computationally simpler (in the sense that it can model a tens of thousands of neuron networks with in a very short time) and is biologically similar to the Hodgin-Huxley model. The Izhkevich neuron model shows a lot of interesting spiking and bursting features and efficient in describing the neuron dynamics. The cortical excitatory neurons are classified into a) reguar spiking b) intrinsically bursting c) chattering where as the cortically inhibitory interneurons are classified into a) fast spiking b) low threshold c) spiking. IZH neuron model shows all of the above spiking and bursting features as the parameters in the system are varied. In addition, the model also describes the behavior of thalamo-cortical neuron, which is one of the major source of input to the cortex. The model can also exhibit a resonator neuron dynamics.

A neuron is said to fire an action potential when its membrane potential exceeds a limiting value. Interpreting this feature in brain, the neuron is excited when there is an chemical input to its synapses. It is found that neurons can also be excited externally using electric field, magnetic field (using electrodes in a bath for input). A careful application of electric field on embroynic neuronal cells have led to their growth in cultures \cite{Ste17}. Therefore it is of interest to consider the effect of external electric, magnetic field, Guassian noise on the dynamical behavior of the neurons. To model it mathematically and in order to improve the existing neuron models and make it very similar and mimic the real life scenarios, researchers have considered the effect of external features on the model (additional terms in the existing models). Adding such external effect can increase the dimension of the neuron model and thereby increasing the chances of exploring diverse regimes and phenomenons and can open doors to rich nonlinear dynamics. For instance, it can show many global bifurcation phenomena, period-doubling cascade route to chaos. Similar effect of external magnetic field is studied in \cite{Ka20}, where the original two dimensional continuous ODE Izhikevich neuron model where in the presence of external magnetic field transforms into a three dimensional continuous ODE model. For instance, the spiking and bursting pattern changes. It is shown in \cite{Ka20} that periodic spiking and bursting are observed in the presence of external magnetic field.  

Recently, researchers have found that discrete maps are also effective and efficient in modelling the neurons as compared to continuous ODE models. Moreover, few ODE models have exact solutions while others need to integrated numerically for which we resort to discretisation. It is easier to iterate a discrete map than to integrate a continuous dynamical system. This is really efficient while modelling a large neural network which is often computationally intensive in the case of continuous system. There are a plethora of discrete neuron models describing the neuron dynamics.  Discretised neuron models include Rulkov map, a discrete two dimensional model exhibiting tonic spiking, chaotic bursts. Many of such discrete maps have become popular recently and have been reviewed in \cite{Ibar11}. 

After understanding the dynamics of individual IZH neuron, we try to understand how a network of IZH neurons behave. The basic structure of the brain is composed of large number of neurons interconnected in a complex fashion. We try to gain some insights about how the neuron behave under different networks. Synchronization is a universal phenomenon in which several elements of the network synchronize and behave similarly in the sense that phase and frequency are similar. But recently, Kuramoto found that networks with synchrony can pass through an intermidiate stage where there is a coexistence of synchronized and desynchronized states \cite{Kur02}. These states are referred to as chimera states. The presence of chimera states, synchronization has effects on neuronal systems and is also thought of as a mechanism leading to several neuronal disorders like epilepsy, schizophernia \cite{Arum15}. It is found that the neuronal diseases are topological in the sense that it depends on the topolgy of the network of neurons interconnected. Hence understanding the dynamical behavior of a network of neurons is important. As discretised IZH neuron model is computationally faster and can be used to simulate large networks, we next consider different networks of IZH neurons. There has been a lot of studies concerning networks of dynamical systems, that is under what conditions they synchronize, desynchronize and the presence of chimera states. Ring and star networks are simple yet most important topologies which are found in many genetic, biological studies and even manifest themselves in many real world networks. Synchronization properties of star connected identical Chua circuits have been studied in \cite{Muni18}. Here, We showcase IZH neurons connected in ring and star networks. Next, we also consider a mixed topological network introduced in \cite{Muni20} which is known as ring-star network and study for the prevalence of chimera states. The IZH neurons in all of the networks are non-locally diffusively coupled.  

In this paper, we consider the discretised version of Izhkevich neuron model in the presence of external electromagnetic field. We discretise it using forward Euler method. The action of external magnetic field is mimiced by adding an extra coupling term. Depending on the sign of the coupling strength, we analyse their dynamics under the action of active external coupling (positive coupling strength), resistive coupling (negative coupling strength). We explore the dynamical behavior of the IZH neuron with the variation of the external coupling strength and the synapses current. The dynamical behavior is guided by the techniques of bifurcation diagrams, Lyapunov exponent, two parameter Lyapunov exponent phase diagrams to explore various dynamical regimes in the parameter space.  

\section{Improved discretised IZH neuron model}
Izhikevich in 1974 described a neuron model as follows:
\begin{equation} \label{eq:IZHorg}
\begin{aligned}
    \dot{v} &= v + 0.04 v^{2} + 5 v +I + 140 - u,\\
    \dot{u} &= u + a (bv - u),
\end{aligned}
\end{equation}
where $u,v$ represent the state variables and describe the potential of the neuron membrane and the recovery variable respectively. The variables $a,b$ are dimensionless parameters. The model above is computationally effective as compared to other neuron models in the sense that it can model a large network of Izhkevich neuron system within seconds.
In this paper we consider an improved  discretised version of Izhkevich neuron model under the action of external electromagnetic field. Here we discretise the system \eqref{eq:IZHorg} under the forward Euler scheme. The improved discretised Izhkevich neuron model is then described as follows:
\begin{equation} \label{eq:IZH_discrete}
\begin{aligned}
    v(i+1) &= v(i) + 0.04 v(i)^{2} + 5 v(i) + 140 + I - u(i) + kv(i)M(\phi),\\
    u(i+1) &= u(i) + a (bv(i+1) - u(i)),\\
    \phi(i+1) &= \phi(i) + k_{1}v(i+1) - k_{2}\phi(i),
\end{aligned}
\end{equation}
The value of $v$ is trapped back if it exceeds a threshold value (here $30$). Then $v,u$ follow the rule as below:
if $(v(i+1)) \geq 30\rm{mV},$\\
 $v(i+1) \leftarrow c$,\\
 $u(i+1) \leftarrow u(i) + d$.\\
end
where $u,v,\phi$ are state variables representing the potential of the neuron membrane, recovery variable and the magnetic flux respectively. The variables $a,b,k_{1},k_{2},c,d$ are dimensionless parameters respectively. The term $M(\phi)$ represents memory conductance and is set as $M(\phi) = \alpha + 3\beta \phi^{2}$, where $\alpha, \beta$ are fixed constants. The variables $k,I$ represent the coupling strength of the external electromagnetic field and are considered as parameters in this study. We analyse the bifurcation diagrams, Lyapunov exponents with respect to the parameters $k, I$. It should be noted that $k$ can take positive (active coupling) and negative (resistive coupling) values. We also explore this effect through the bifurcation diagram.

\begin{figure}[!htbp]
\begin{center}
\includegraphics[width=0.7\textwidth]{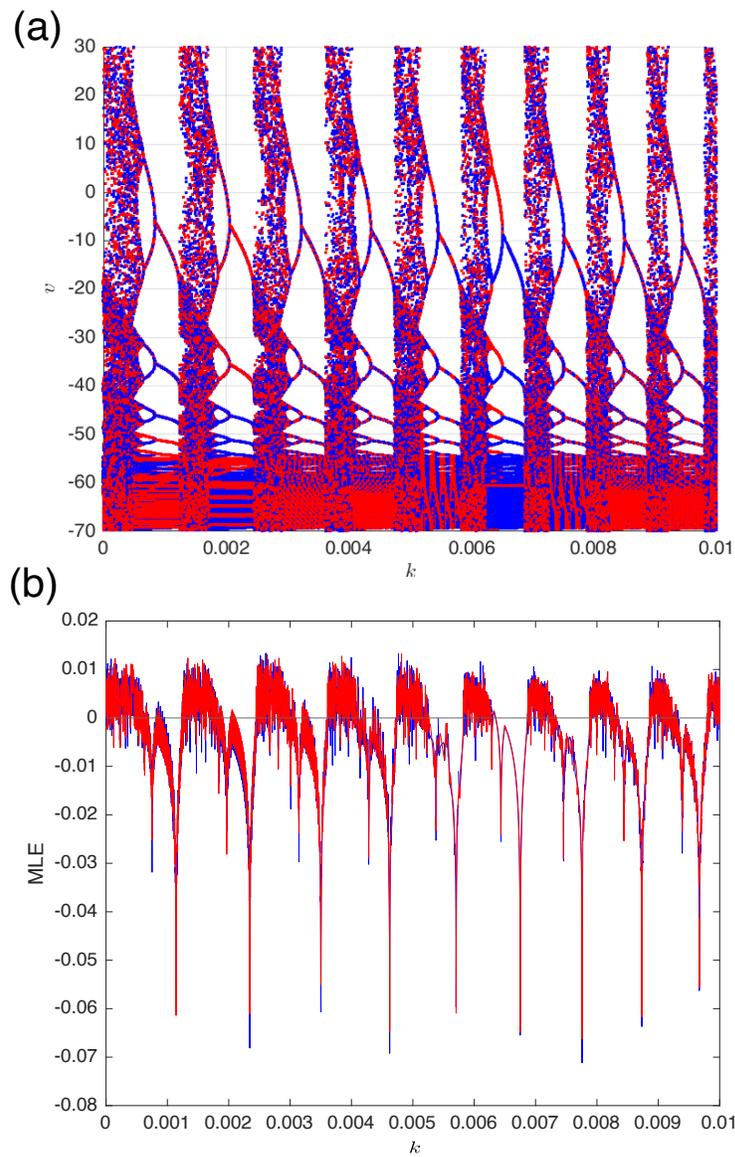}
\end{center}
\caption{Blue: Forward continuation, Red: backward continuation bifurcation diagram of $v$ vs $k$ for $I = 1$. The case for active coupling $k > 0$. Blue: Forward continuation, Red: backward maximal Lyapunov exponent diagram of $v$ vs $k$ for $I = 1$.}
\label{fig:bif_vk}
\end{figure}

The improved Izhkevich neuron model shows chaotic behavior in contrast to the original model which is a two dimensional continuous system in which there is no chaos (from Poincar\'e-Bendixson theorem). The effect of introducing external electromagnetic field is studied in this paper.
The Parameters are fixed as  $a = 0.02, b = 0.25, c = -55, d = 2, I = 1,
k = 0.01, k_{1} = 0.01, k_{2} = 0.1, \alpha = 0.1, \beta = 0.001.$ 
Figure \ref{fig:bif_vk} shows the forward and backward continuation bifurcation diagrams with respect to different state variables and parameters. We observe multistability here as the forward and backward bifurcation diagrams do not overlap completely and hence they are distinct and represent two different dynamical states when forward and backward direction in the $k$ parameter space is considered. 

In Figure\ref{fig:bif_vk}(a), we observe that variation with $k$ shows period-doubling route to chaos in a repeating fashion. The blue points correspond to the forward continuation that is value of $k$ is increased from $0$ to $0.01$ with new initial conditions taken for each further $k$ value as the last data point of the state variables. Similarly, the red point correspond to the backward continuation that is the value of $k$ is decreased from $0.01$ to $0$ with new fresh intiial conditions taken for each subsequent $k$ value as the last data point of the state variable. This is also confirmed with the help of Figure \ref{fig:bif_vk}(b), in which we can see the chaotic regions corresponding to the positive values of the Lyapunov exponents.

\begin{figure}[!htbp]%
    \centering
   \hspace*{-0.5cm}\includegraphics[width=0.7\textwidth]{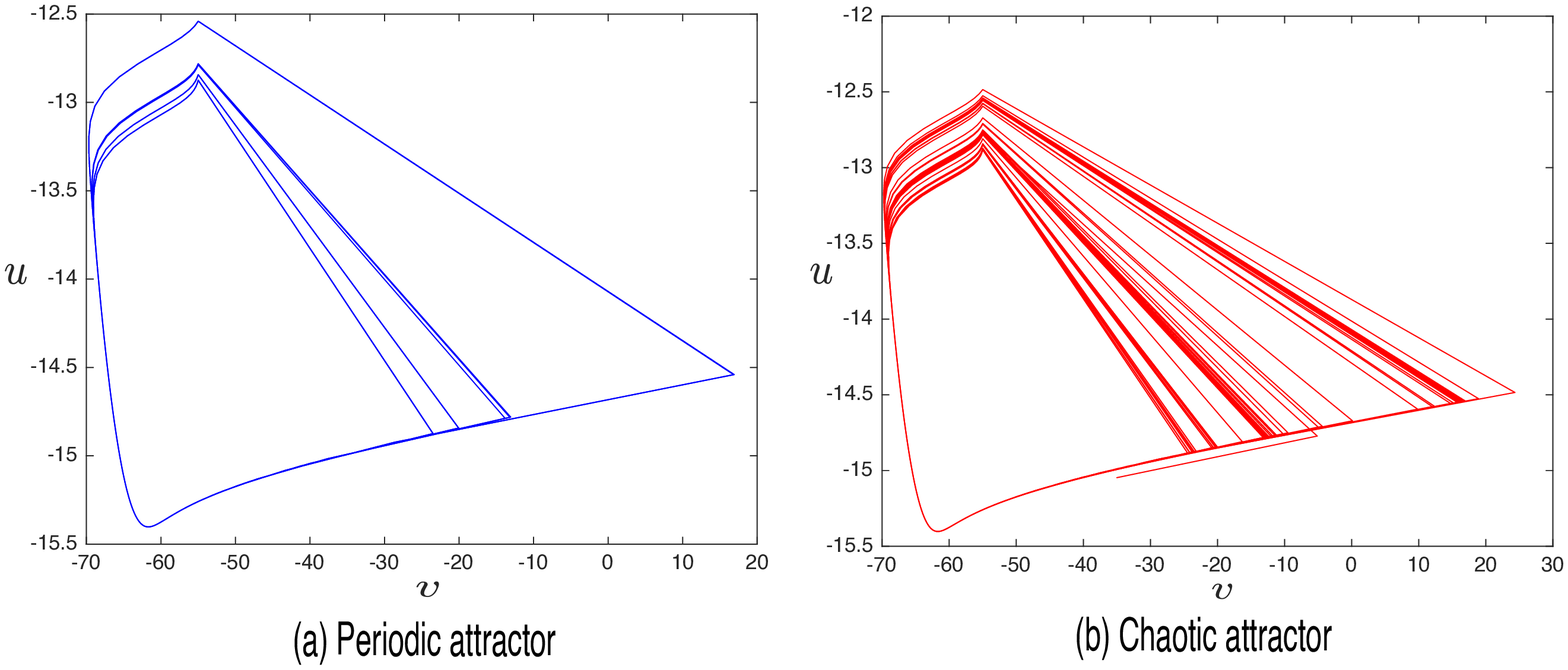}
    \caption{Coexistence of periodic and chaotic attractor at $k=1.280256$. Periodic attractor is shown in (a) for forward continuation in blue colour and chaotic attractor is shown in (b) for backward continuation in red colour. The choice of $k$ value for the coexistence is deduced from figure \ref{fig:bif_vk}.}%
    \label{fig:coexistPosK}%
\end{figure}

In figure \ref{fig:coexistPosK}, we observe the coexistence of periodic and chaotic attractors for different choice of initial conditions and for a fixed value of $k =1.280256$. Periodic attractor in blue is observed for initial condition as $(-6.5878960e+01,-1.4912318e+01,-6.6153338e+00)$ and chaotic attractor is observed for initial condition as $(-6.2159677e+01,-1.5491414e+01,-6.3154825e+00)$. This shows bistability in the discretised IZH model for $k>0$ region.

For the same range of the parameter $k$, we analyse the variation of maximal Lyapunov exponent computed in both forward and backward directions. The Lyapunov exponent is computed using the \textit{QR} factorisation method. 
We observe that chaotic regions have positive maximal Lyapunov exponent. 

\begin{figure}[!htbp]
\begin{center}
\includegraphics[width=0.7\textwidth]{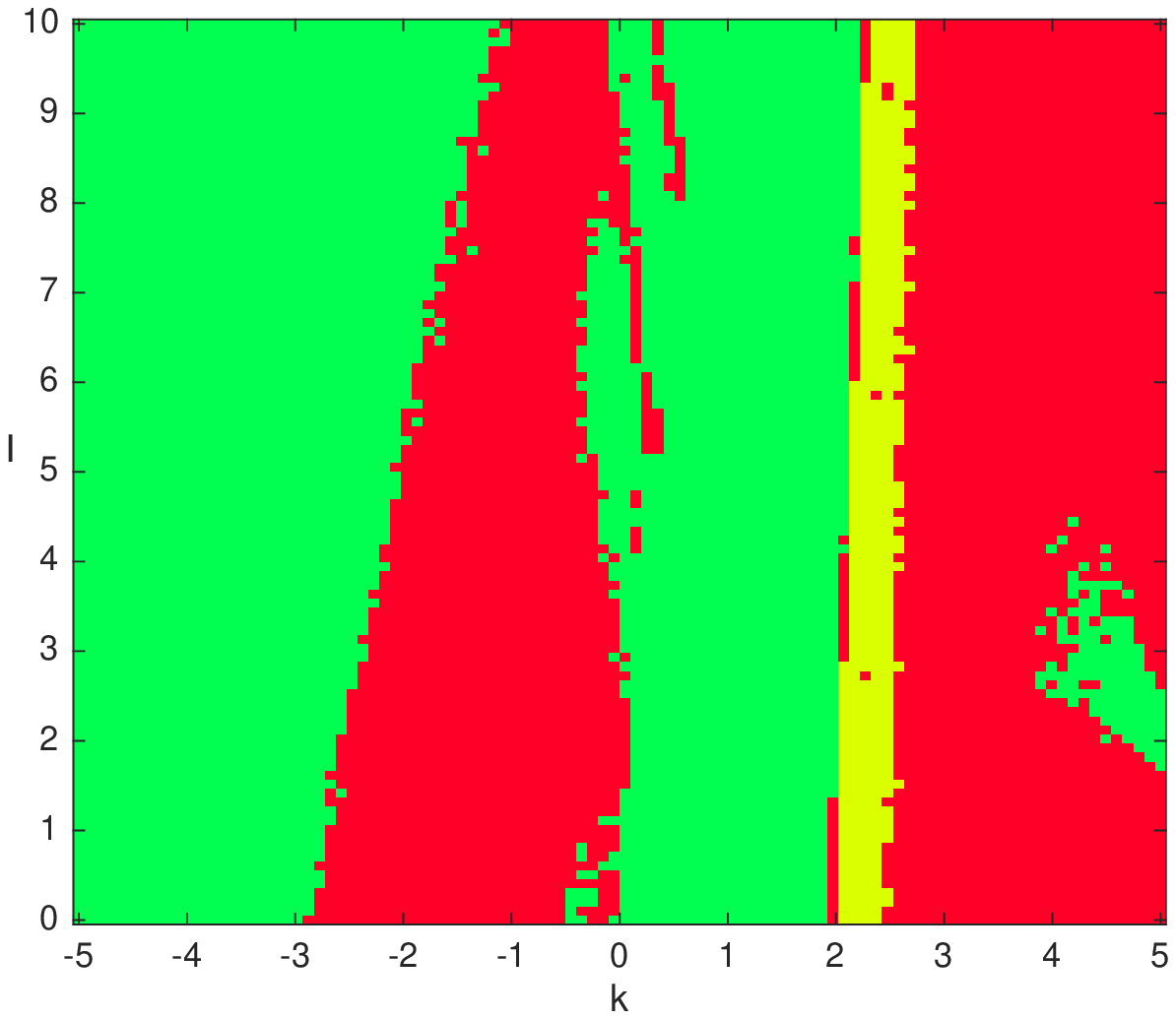}
\end{center}
\caption{ Lyapunov $k-I$ basin diagram. Red colour represents chaotic region, yellow colour represents unbounded region, green colour represents periodic region.}
\label{fig:LEbasinposk}
\end{figure}

Next in Figure \ref{fig:LEbasinposk}, we consider the two parameter $K-I$ Lyapunov exponent basin plot. This can provide us with the dynamical behavior in the $K-I$ parameter space like chaotic, periodic, diverging behavior. We find that there is a presence of large area of periodic and chaotic behavior with a strip of unbounded regions. Even the chaotic regions in red appear to converge to a shrimp shaped domain. 
\section{Spiking and bursting patterns}
\begin{figure}[!htbp]
\begin{center}
\includegraphics[width=1.1\textwidth]{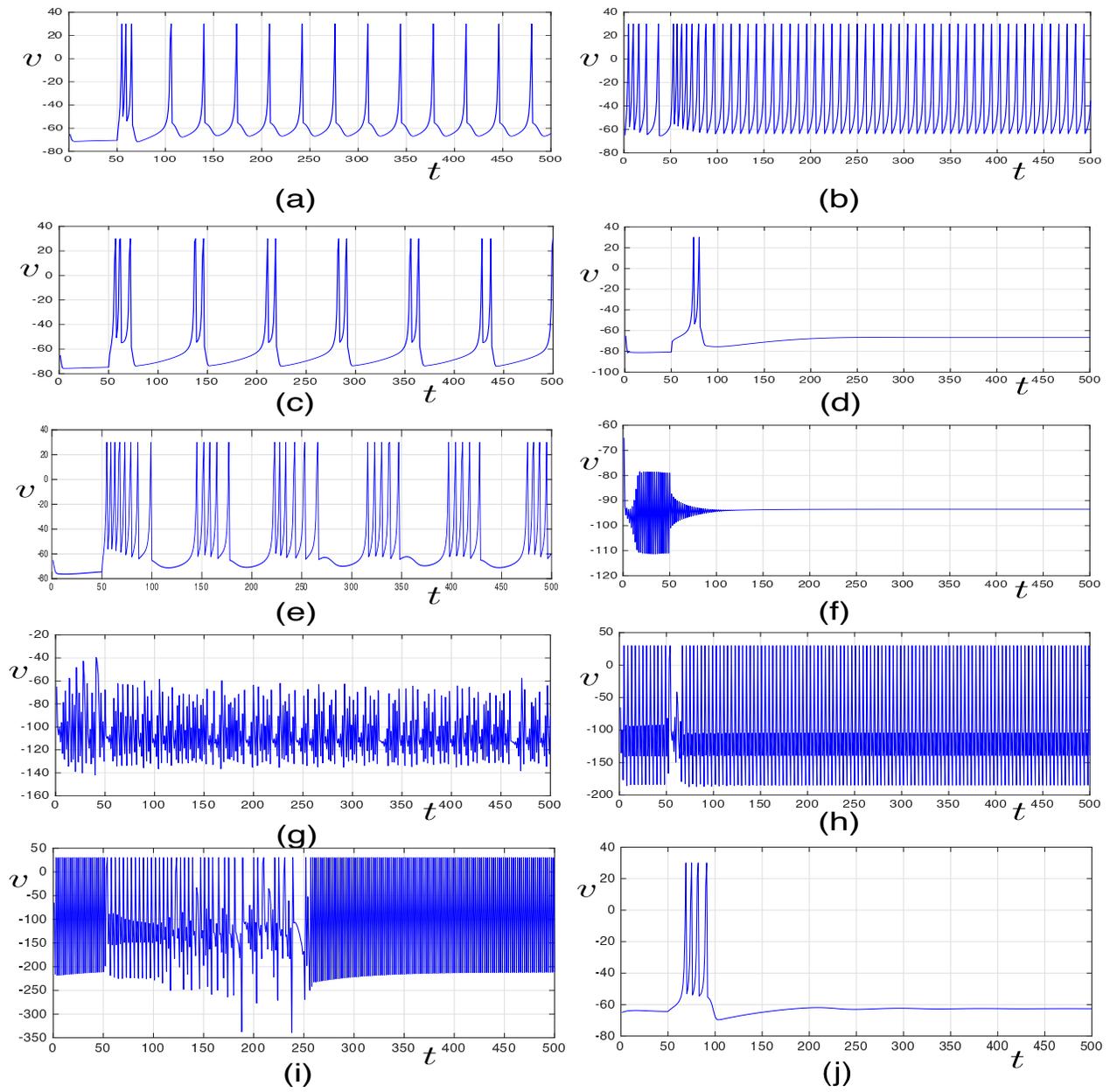}
\end{center}
\caption{Various firing patterns in the IZH model with the variation of parameters $k$ and $I$. The parameters set as  ($a = 0.02, b = 0.4, c = -65, d = 2, I =10$).}
\label{fig:FiringPatternIZH}
\end{figure}

Here we illustrate a plethora of different spiking and bursting patterns shown by the discretised IZH model \eqref{eq:IZH_discrete}. We consider the case when there is no external electromagnetic field and then we consider the effect of the electromagnetic field on the spiking and bursting pattern as the electromagnetic coupling $k$ is varied.
 For the set of parameter values, tonic spiking is observed. For nearby coupling strength $k = 0$, the tonic spiking persists. When the value of $k$ increases to $0.9$, periodic bursts are observed as in figure \ref{fig:FiringPatternIZH}(e). When the magnetic field coupling strength increases to $2.8$, chaotic spiking is observed (see figure \ref{fig:FiringPatternIZH}(g)). Further increase in the value of $k$, leads to a regular spiking pattern as in figure \ref{fig:FiringPatternIZH}(h). A large increase in the value of $k  = 10$ brings back the chaotic spikes with some regularities initially as in figure \ref{fig:FiringPatternIZH}(i). With an increase in the synapses current in figure \ref{fig:FiringPatternIZH}(a), we observe a mixed mode pattern. With further increase in the electromagnetic flux, we observe the spike bursting in figure \ref{fig:FiringPatternIZH}(c) and \ref{fig:FiringPatternIZH}(f) and further increase in $k$, we observe chaotic spiking in figure \ref{fig:FiringPatternIZH}(d). The presence of tonic spiking in figure \ref{fig:FiringPatternIZH}(b) for $k=10,I=10$. In the absence of external electromagnetic field $(k=0)$ and  a reduced value of synapses current $I = 0.6$, we observe phasic bursting pattern in figure \ref{fig:FiringPatternIZH}(j).

\section{Resistive coupling}
\begin{figure}[!htbp]
\begin{center}
\includegraphics[width=0.7\textwidth]{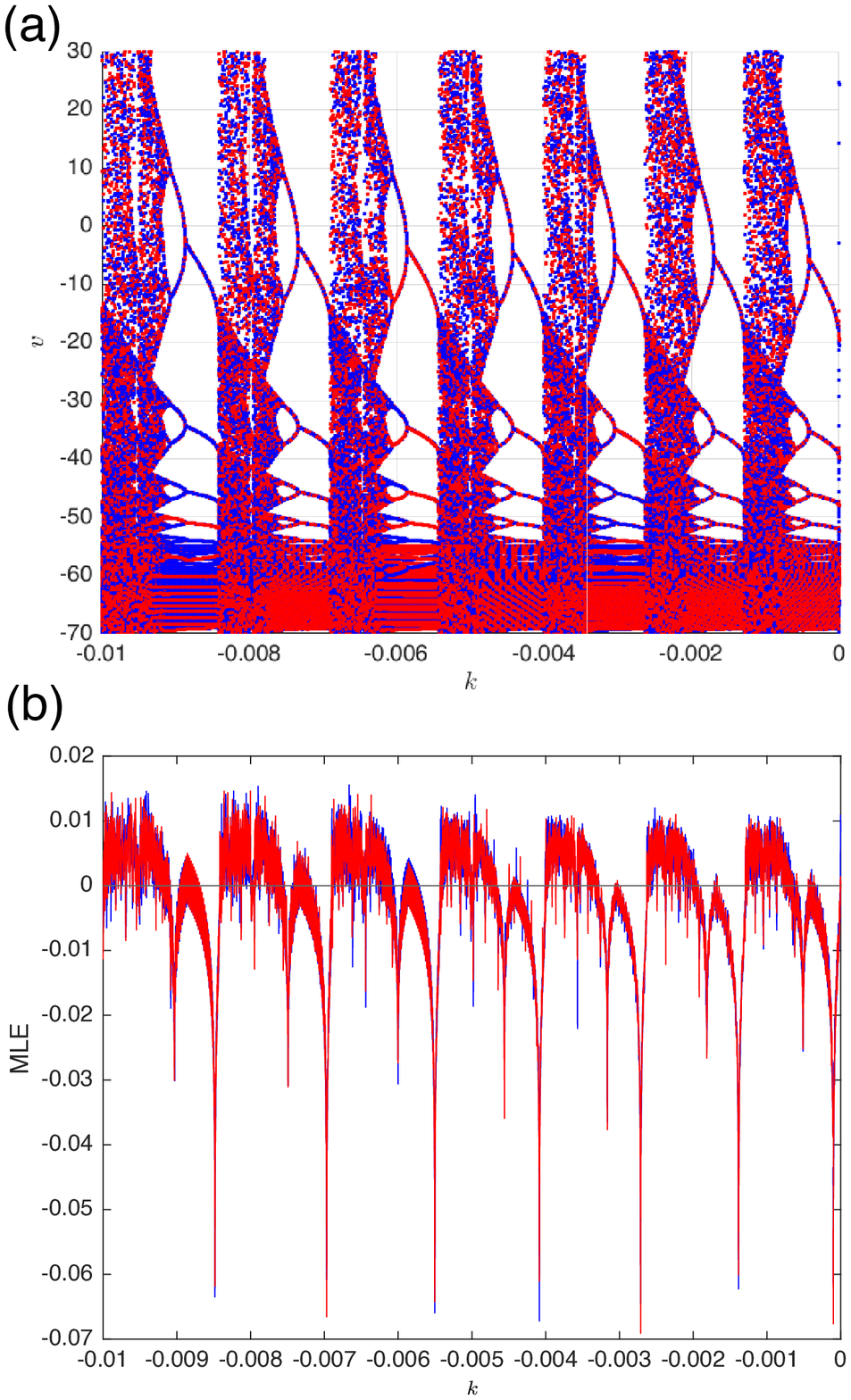}
\end{center}
\caption{Blue: Forward continuation, Red: backward continuation bifurcation diagram of $v$ vs $k$ for $I = 1$. The case for resistive coupling $k < 0$. Blue: Forward continuation, Red: backward maximal Lyapunov exponent diagram of $v$ vs $k$ for $I =1$.}
\label{fig:bif_vk_kneg}
\end{figure}

The external electromagnetic field  is resistively coupled when $k<0$. Similar bifurcation structure is obtained for $k < 0$. In Figure \ref{fig:bif_vk_kneg}(a), we observe that variation with $k$ we observe period-doubling route to chaos in a repeating fashion. The blue points correspond to the forward continuation that is value of $k$ is increased from $-0.01$ to $0$ with new initial conditions taken for each further $k$ value as the last data point of the state variables. Similarly, the red point correspond to the backward continuation that is the value of $k$ is decreased from $-0.01$ to $0$ with new fresh initial conditions taken for each subsequent $k$ value as the last data point of the state variable. This is also confirmed with the help of Figure \ref{fig:bif_vk_kneg}(b), in which we can see the chaotic regions corresponding to the positive values of the Lyapunov exponents.  We observe period-doubling bifurcation route to chaos repeating itself as the parameter $k$ varies.

We also observe multistability for the resistive coupling region as evident from figure \ref{fig:bif_vk_kneg}.
\begin{figure}%
    \centering
    \includegraphics[width=1\textwidth]{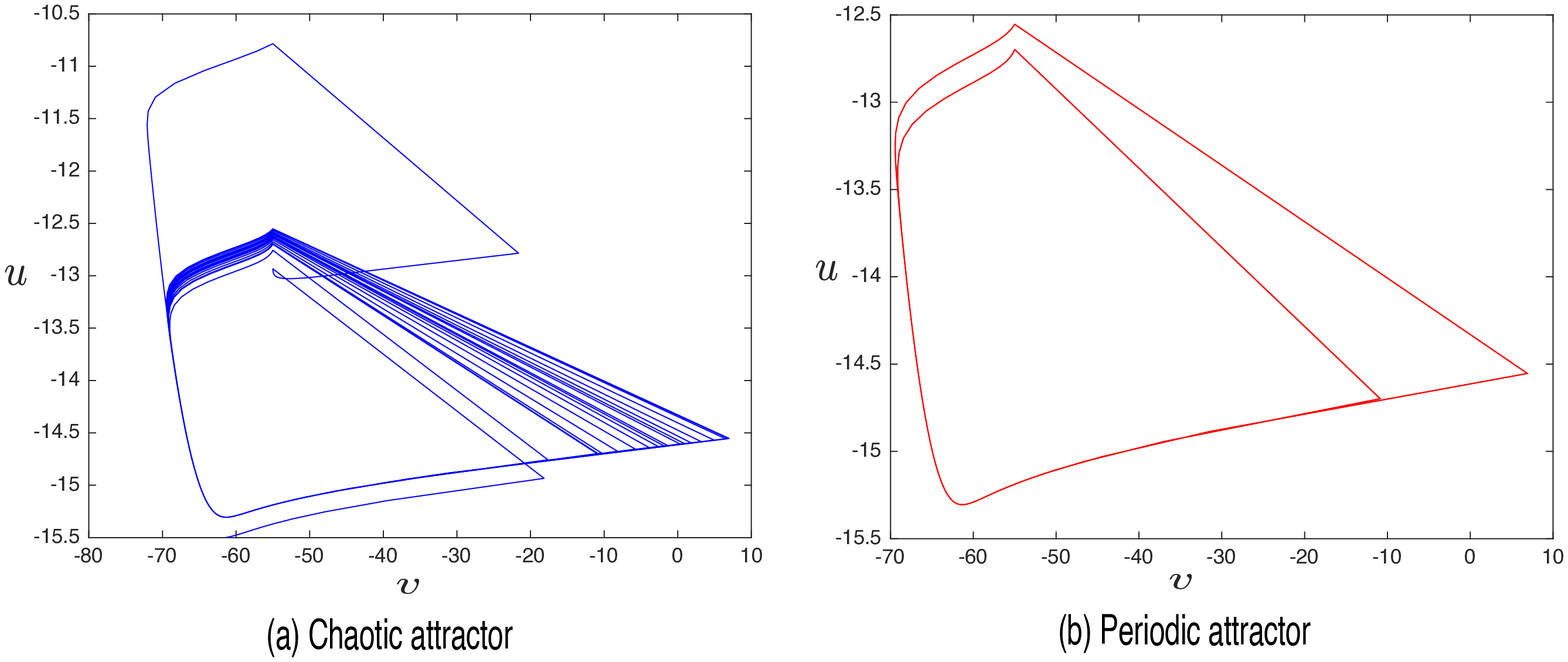}%
    \caption{Coexistence of periodic and chaotic attractor at $k=-0.009$ for different initial conditions. }%
    \label{fig:coexistNegK}%
\end{figure}
In figure \ref{fig:coexistNegK}, we observe the coexistence of periodic and chaotic attractors for different choice of initial conditions and value of $k$ fixed as $k=-0.009$. Periodic attractor in blue is observed for initial condition as $(-6.135426e+01,-1.549789e+01,-6.246161e+00)$ and chaotic attractor is observed for initial condition as $(-6.441688e+01,-1.509566e+01,-6.526226e+00)$. This shows bistability in the discretised IZH model for $k<0$ region.

\section{Chimera state in IZH networks}
\begin{figure}[h!]
\begin{center}
\includegraphics[width=0.5\textwidth]{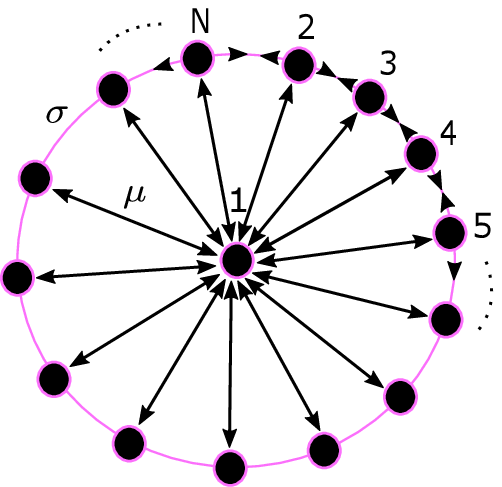}
\end{center}
\caption{The ring-star network of discrete IZH neurons. Here, we consider $N = 100$ IZH neuron system where the central node is labelled as $m=1$ and the end nodes are labelled from $m= 1, …N$. The star coupling strength from the central node $m=1$ to the end nodes $m=2,\ldots,N$ is denoted by $\mu$. The ring coupling strength between the end nodes is denoted by $\sigma$.}
\label{fig:ringstar}
\end{figure}

Investigating the network behaviour of neurons plays a significant role in understanding the synchronous and asynchronous regimes in the coupled neuron network \cite{Arum15,Maj17,Mak16}. Many literatures have shown that the coexistence of both synchronous and asynchronous behaviours plays a vital role in studying the brain neuronal activity \cite{Khal19}. Though there have been many literatures discussing the existence of chimeras in differential equation type Izhikevich neurons \cite{Sha19,Ka20} but less has been the investigation on the map based Izhikevich models. Also (additionally), we have considered magnetic flux coupling in the discrete Izhikevich (IZH model) to study the effect of magnetic field in the neuron dynamics. We have considered a ring-star topology as in Fig.\ref{fig:ringstar} for analysing the network dynamics of the discrete IZH neurons. Each node $m = 1,2, …, N$ denote the discrete IZH neuron system whose dynamics follow \ref{eqn:RingStarIZHEqn}.
The mathematical model of the ring-star connected discrete IZH neuron map is defined as,
\begin{equation}
    \begin{aligned}
     v_{m}(i+1) &= v_{m}(i) + 0.04 v_{m}(i)^{2} + 5 v_{m}(i) + 140 - u_{m}(i) + I + kv_{m}(i)M(\phi_{m}(i)) \\ +& \mu (v_{i} - v_{m}(i)) + \frac{\sigma}{2P} \sum_{n=m-P}^{m+P} (v_{n} - v_{m}),\\    
    u_{m}(i+1) &= u_{m}(i) + a(bv_{m}(i+1) - u_{m}(i)),\\
    \phi_{m}(i+1) &= \phi_{m}(i) + k_{1}v_{m}(i) -k_{2}\phi_{m}(i)
    \end{aligned}
    \label{eqn:RingStarIZHEqn}
\end{equation}

whose central node is defined as
\begin{equation}
    \begin{aligned}
    v_{1}(i+1) &= v_{1}(i) + 0.04 v_{1}(i)^2 + 5v_{1}(i) + 140 - u_{1}(i) + I + kv_{1}(i)M(\phi_{1}(i)) + \mu \sum_{n=1}^{N}(v_{n}-v_{1}),\\
    u_{1}(i+1) &= u_{1}(i) + a(bv_{1}(i+1) - u_{1}(i)),\\
    \phi_{1}(i+1) &= \phi_{1}(i) + k_{1}v_{1}(i+1) - k_{2}\phi_{1}(i)
    \end{aligned}
    \label{eqn:CentRingStarEqn}
\end{equation}

with periodic boundary conditions:
\begin{equation}
    \begin{aligned}
    u_{i+N}(t) &= u_{i}(t),\\
    v_{i+N}(t) &= v_{i}(t),\\
   \phi_{i+N}(t) &= \phi_{i}(t).
    \end{aligned}
    \label{eqn:BC}
\end{equation}

where the reset in state variables is as below
If $v \geq 30$mV, then
\begin{equation*}
    \begin{aligned}
    v &\leftarrow c,\\
    u &\leftarrow u + d.
    \end{aligned}
\end{equation*}

 The star coupling strength denoted by $\mu$ is the coupling strength between the neurons in the star network and with the central node, $\sigma$ is the ring coupling strength between the neurons in the ring network, $P$ denotes the number of neighbours. The magnetic induction is modelled by the well-known cubic memductance function  $M(\phi_{m}(i)) = \alpha + 3 \beta \phi_{m}(i)^2$. 
 
 The network size is defined as $m=1$ to $N$ and for simulations we took $N=100$, $P=10$ while the other system parameters are fixed to their respective values. We have divided the entire analysis into three categories where in category A we discussed a simple ring network by considering $\mu=0$ and considered $\sigma$ as the control variable and in category B we investigated ring-star network by considering both $\mu$ and $\sigma$ as the variables of discussion. In the final category C we discussed star connected network with $\sigma = 0$. In all the discussions we have used recurrence plots using the mathematical analysis described by  $R_P= ||v_i-v_j ||$ for $i,j=[1,2,3,...,N]$, where $||.||$ denotes the Euclidean norm.

\subsection{Category-A (Ring network)} 
\begin{figure}[h!]
\begin{center}
\includegraphics[width=0.7\textwidth]{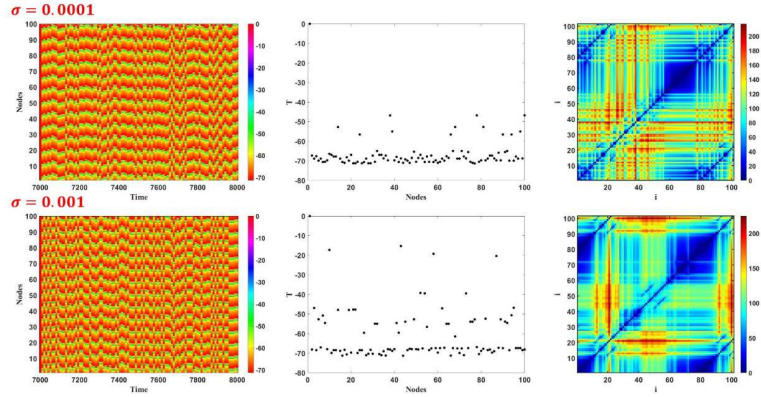}
\end{center}
\caption{All the nodes in the ring network in asynchronous state for different coupling strength. The central node is eliminated from the network topology by fixing $\mu=0$. Random initial conditions are used to increase the complexity of the nodes connected. The left most plot shows the spatiotemporal dynamics, the middle plot shows the end state values of the nodes and the right plot shows the recurrence plot of the nodes.}
\label{fig:RingChim1}
\end{figure}

\begin{figure}[htbp!]
\begin{center}
\includegraphics[width=0.7\textwidth]{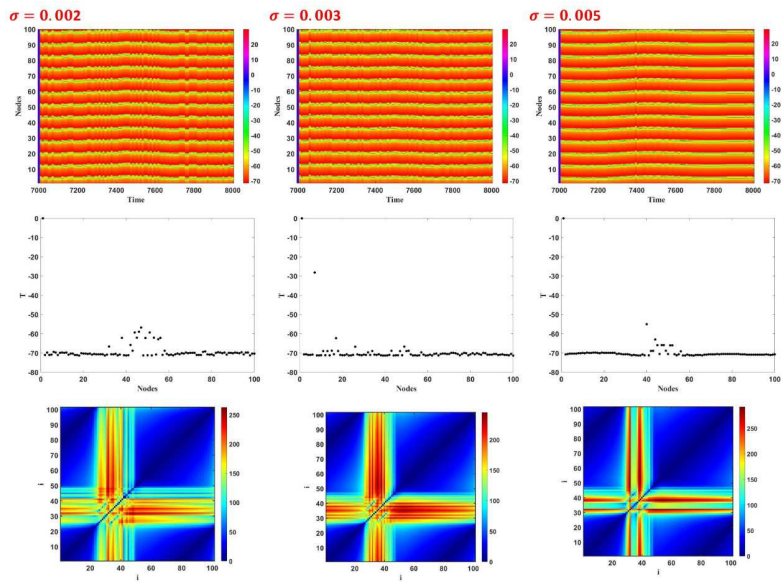}
\end{center}
\caption{All the nodes in the ring network in chimera state for different coupling strength. The central node is eliminated from the network topology by fixing $\mu=0$. Random initial conditions are used to increase the complexity of the nodes connected. The top most plot shows the spatiotemporal dynamics of the neurons in the ring network, the middle plot shows the end state values of the nodes and the lower plot shows the recurrence plot of the nodes. We could note the emergence of chimera states from the middle plot with several asynchronous nodes coexisting with the synchronous nodes seen in the network.}
\label{fig:RingChim2}
\end{figure}

In this section we consider the neurons in a ring connected network alone by using $\mu=0$ and eliminating the central node. For the numerical investigation we have considered the coupling strength $(\sigma)$ of the nodes in the network as the control parameter. The spatiotemporal dynamics of the network considering lower coupling strengths of $\sigma=0.0001$ and $\sigma=0.001$ are shown in Fig.\ref{fig:RingChim1}. While comparing the spatiotemporal plots (left most in Fig. \ref{fig:RingChim1}) with the final state plots (middle plots in Fig.\ref{fig:RingChim1}) we could confirm that all the nodes are in asynchronous state. We have used the recurrence plots (right most in Fig.\ref{fig:RingChim1}) where incoherence is shown by formation of irregular structures. 
When the coupling strength is increased to $\sigma=0.002$, we could note synchronous and asynchronous nodes in the network. Such coexistence of synchronous and asynchronous behaviour in the network is termed as chimeras. Such chimeras can be determined by the formation of regular structures in the recurrence plot shown in Fig.\ref{fig:RingChim2}. The areas shown in blue correspond to the synchronous nodes while those in red show asynchronous nodes. 
\subsection{Category-B (Ring-Star network)}
\begin{figure}[h!]
\begin{center}
\includegraphics[width=0.7\textwidth]{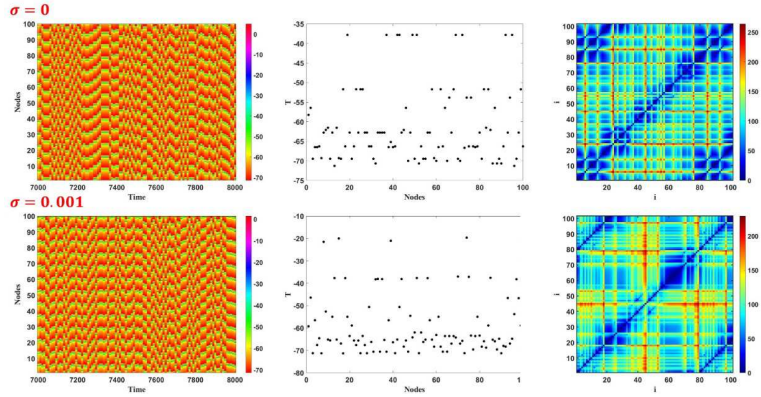}
\end{center}
\caption{All the nodes in the ring-star network are in asynchronous state for different coupling strength. The coupling strength between the central node and the nodes in the ring structure is $\mu=0.005$. Random initial conditions are used to increase the complexity of the nodes connected. The left most plot shows the spatiotemporal dynamics of the neurons in the ring-star network, the middle plot shows the end state values of the nodes and the right most plot shows the recurrence plot of the nodes.}
\label{fig:RingStarChim1}
\end{figure}

In this section we consider a complete ring–star network by fixing $\sigma \neq 0$ and $\mu \neq 0$. To make the discussion simple we fix the value of the star coupling strength between the central node and the nodes in the ring as $\mu = 0.005$. Now we have considered $(\sigma)$ the coupling strength of the nodes in the ring network as the control parameter. For  $\sigma =0$ and $\sigma = 0.001$  the nodes in the network are asynchronous as displayed in fig. \ref{fig:RingStarChim1}. The state plots of the nodes captured at the end of simulation are shown in fig.4 (middle) confirms asynchronous states. The absence of regular structure in the recurrence plots also confirms the incoherent nodes of the network. 

Increasing the coupling strength to $\sigma = 0.002$, the nodes tend to synchronize and a regular structure (blue) in the recurrence plot shows the synchronous nodes as in fig.5. Further increasing to $\sigma = 0.005$  , the network shows both synchronous and asynchronous nodes confirming the emergence of chimera states. These chimera states can be easily verified by recurrence plots when the blue colour denotes synchronous nodes and the other colour denotes asynchronous node.

\begin{figure}[h!]
\begin{center}
\includegraphics[width=0.7\textwidth]{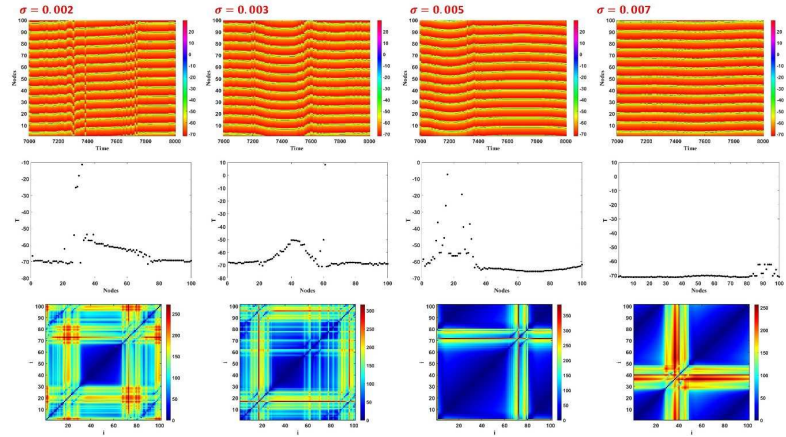}
\end{center}
\caption{All the nodes in the ring-star network in chimera state for different coupling strength. The coupling strength between the central node and the nodes in the ring structure is $\mu=0.005$. Random initial conditions are used to increase the complexity of the nodes connected. The top most plot shows the spatiotemporal dynamics of the neurons in the ring-star network, the middle plot shows the end state values of the nodes and the lower most plot shows the recurrence plot of the nodes. }
\label{fig:RingStarChim5}
\end{figure}

\subsection{Category-C (Star network)}
In the final discussion of this section we consider the star connected network by keeping the coupling strength between the nodes in the ring as $\sigma=0$. In fig.\ref{fig:RingStarChim1} we have shown the spatiotemporal plots (right), the end state values of each node (middle) and the recurrence plots (left) for $\mu=0$ and $\mu=0.001$. The plots confirm that the nodes are in asynchronous states and the recurrence plots show no defined structures.
\begin{figure}[h!]
\begin{center}
\includegraphics[width=0.7\textwidth]{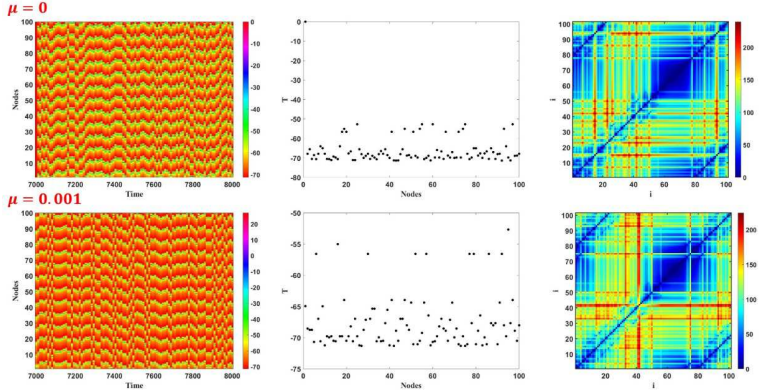}
\end{center}
\caption{All the nodes in the star network are in incoherent state for different star coupling strengths $\mu$. The coupling strength between the nodes in the ring structure is $\sigma=0$. Random initial conditions are used to increase the complexity of the nodes connected. The left most plot shows the spatiotemporal dynamics of the neurons in the star network, the middle plot shows the end state values of the nodes and the right plot shows the recurrence plot of the nodes.}
\label{fig:StarChim1}
\end{figure}
By increasing the value of $\mu$ to $0.003$, the nodes form different local clusters of synchronisations as shown in Fig.\ref{fig:rINGStarChim2}. These cluster can be confirmed by formation of different small structures in the recurrence plots confirming smaller clusters of synchronised nodes. By further increasing the value of $\mu$ to $0.005$ the smaller clusters merge synchronises with their nearest larger cluster and thus forms larger clusters of synchronised nodes. Increasing the value of $\mu$ to $0.008$ doesn’t show any change in the cluster synchronisation. Thus we could confirm that in a star connected network we couldn’t identify the chimera states and only cluster synchronisation of nodes are displayed.
\begin{figure}[h!]
\begin{center}
\includegraphics[width=0.7\textwidth]{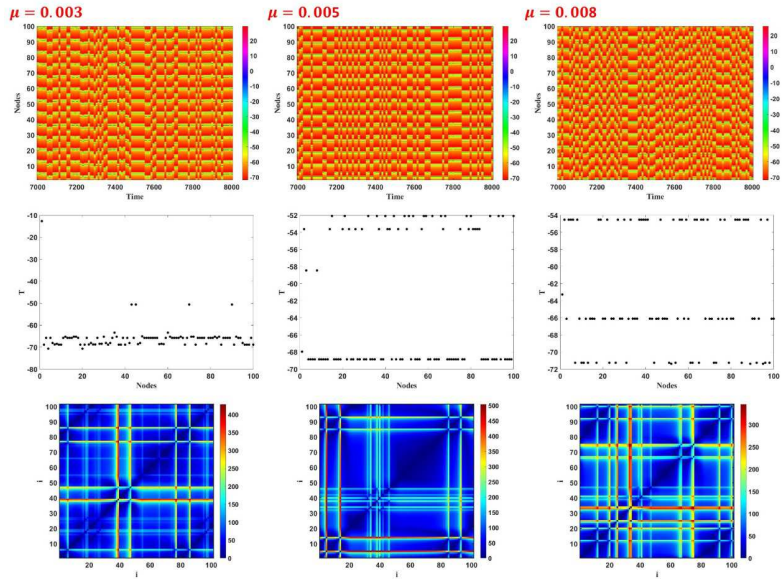}
\end{center}
\caption{All the nodes in the star network in incoherent state for different coupling strength. The coupling strength between the nodes in the ring structure is $\sigma=0$. Random initial conditions are used to increase the complexity of the nodes connected. The upper plot shows the spatiotemporal dynamics of the neurons in the star network, the middle plot shows the end state values of the nodes and the lower plot shows the recurrence plot of the nodes.}
\label{fig:rINGStarChim2}
\end{figure}
\section{Conclusions}
It is found that discretised Izhkevich neuron model exhibits a lot of different bursting and spiking features. Similar to the continuous case, it exhibits regular spiking, intrinsically bursting, chattering, fast spiking, low-threshold spiking. Tuning the magnetic field, it is found that it exhibits the period-doubling cascade to chaos. This is also aided by the plot of maximal Lyapunov exponent. The discretised IZH model shows bistability (coexistence of periodic and chaotic attractors). After analysing a single IZH neuron model, we analyse the networks of IZH neurons. We considered a mixed topological network known as ring-star network. We showcase the presence of chimera states in the ring-star netowrk and ring network. Multiple synchronised clusters were observed in the case of star networks  with the variation of star coupling strength. The detection of the coherence, incoherence, chimera states, multi-cluster synchronised nodes were aided by the spatiotemporal plots and recurrence plots. 

\section*{Acknowledgements}
This work is partially funded by Centre for Nonlinear Systems, Chennai Institute of Technology, India vide funding number CIT/CNS/2021/RD/064.

\bibliographystyle{elsarticle-num} 

\end{document}